\documentclass[twocolumn,showpacs,showkeys,prb]{revtex4}
\usepackage{graphicx}
\usepackage{dcolumn}
\usepackage{bm}
\usepackage{umlaut}
\begin{document}
\preprint{PREPRINT}

\title{Fermi-surface mapping from Compton profiles: Application to beryllium}

\author{G. Kontrym-Sznajd and M. Samsel-Czekala}
\affiliation{W.Trzebiatowski Institute of Low Temperature and Structure
Research, Polish Academy of Sciences, P.O. Box 1410, 50-950
Wroclaw 2, Poland}
\author{S. Huotari, K. H{\"a}m{\"a}l{\"a}inen, and S. Manninen}
\affiliation{Division of X-ray Physics, Department of Physical Sciences,
P.O. Box 64, FIN-00014 University of Helsinki, Finland}

\date{\today}

\begin{abstract}

The two-dimensional momentum density of Be on the basal $\Gamma
MK$ plane, i.e.~the line integral of the three-dimensional
momentum density along the $c$-axis, is reconstructed via the
Cormack method from both experimental and theoretical Compton
profiles. It is shown that in Be, despite the momentum density is
highly anisotropic, merely two Compton profiles are sufficient to
reproduce the main features of the momentum density. The analysis
of the reconstructed densities is performed both in the extended
and reduced zone schemes.

\end{abstract}

\maketitle

\section{Introduction}

During the last few decades, Compton scattering has been widely
used as a probe of the electronic structure of matter. Within
certain approximations \cite{cooper85,manninen00}, the experimental
double-differential inelastic x-ray scattering cross section can
be related to the Compton profile, which is a double (plane)
integral of the electron density in the extended momentum space,
$\rho(\mathbf{p})$,
\begin{equation}
\label{e1}
J(p_z)=\int_{-\infty}^\infty ~ \rho({\bf p}) ~ {\rm d} p_x {\rm d} p_y.
\end{equation}
The Compton profile, being related directly to
occupied electronic ground states,
contains information not only on the Fermi surface
(FS) but also on the Umklapp components of the electron wave functions.

Eq.~(\ref{e1}) represents the Radon transform \cite{ref2} of
$\rho({\bf p})$ in three-dimensional (3D) space over 2D
hyperplanes. The first reconstruction method based on the
inversion of the Radon transform was Mijnarends' method
\cite{ref3}. Other alternative solutions are Fourier transform
algorithms \cite{ref4,hansen87,ref5,tanaka93}, the Maximum Entropy technique
\cite{ref6} and two different methods employing orthogonal
polynomials \cite{ref7,ref8}.

In cases where the electron momentum density is a strongly
anisotropic function, reconstruction of the 3D momentum density
from 1D Compton profiles is rather elaborate, since a large number
of measurements is needed.
Thus a reconstruction of 2D
momentum densities from Compton profiles \cite{ref11,grazyna93} is proposed, i.e. a
conversion from double (plane) integrals in Eq.~(\ref{e1}) to a
single (line) integral  of the form
\begin{equation}
\rho^{\bf L}(p_z,p_y) \equiv
J(p_z,p_y)= \int_{-\infty}^\infty \rho({\bf p})~ {\rm d} p_x,
\end{equation}
where ${\bf L} || p_x$. When studying the structure of the Fermi
surface, this 2D momentum density can be interpreted much more
easily than the Compton profiles. Furthermore, it can be directly
compared to corresponding 2D-ACAR (two-dimensional angular
correlation of positron annihilation radiation) results, yielding
additional information on the electron-positron interaction. It is
also much easier to reconstruct accurately than the 3D momentum
density $\rho(\mathbf{p})$ because the number of required Compton
profiles is relatively small. 2D momentum densities can also be
folded into the first Brillouin zone (BZ) in order to delineate
various FS elements.

This procedure is demonstrated for beryllium metal by using two
Compton profiles, taken with the $p_z$-axis along the reciprocal
lattice vectors [100] and [110]. In this case
the 2D momentum density can be reconstructed
with the integration axis along [001]. The
resulting densities are shown both in the extended and reduced
zone schemes. For obtaining the density in the reduced zone
scheme, the Lock-Crisp-West (LCW) procedure \cite{ref9} is used.

\section{Conversion from 1D to 2D spectra}

The purpose of this work is to reconstruct $\rho^{\bf
L}(p_{z},p_{y})$, with ${\bf L} || [001]$, from its line
integrals, i.e.~from the Compton profiles

\begin{equation}
\label{e16}
J(p_{z})=\int^{\infty}_{-\infty}\rho^{\bf L}(p_{z},p_{y}) {\rm d} p_{y}.
\end{equation}
For this, the Cormack method \cite{ref10,cormack64,grazyna90} is proposed,
corresponding to the solution of the Radon
transform in 2D space over 1D hyperplanes.

If the direction $p_x$ is along an $R$-fold rotation axis of the
lattice, both the Compton profiles and the 2D momentum density can
be expanded into a cosine series
\begin{equation}
\label{e17}
\rho^{\bf L}(p_{z},p_{y})\equiv\rho(p,\Theta)=
\sum_{i=0}^{\infty}\rho_{iR}^{\bf L}
(p)\cos(iR\Theta)
\end{equation}
\begin{equation}
\label{e18} J(p_{z})\equiv
J(t,\varphi)=\sum_{i=0}^{\infty}g_{iR}(t) \cos(iR\varphi).
\end{equation}
Here $(t,\varphi)$, where $t=|p_z|$, are the polar coordinates of
$p_z$ in the laboratory coordinate system, while $(p,\Theta)$ are
the polar coordinates of the momentum ($p_z,p_y$) in the crystal
coordinate system.

By applying the Cormack method \cite{ref10}, $\rho^{\mathbf
L}(p_z,p_y)$ can be evaluated from Eq.~(\ref{e17}) where its
radial components $\rho_l^{\mathbf L}(p)$ for $l=iR$ are given by
\begin{equation}
\label{e19}
\rho_{l}^{\mathbf L}(p)=\sum_{k=0}^{\infty}(l+2k+1)a_{lk}R^{k}_{l}(p).
\end{equation}
$R_{l}^{k}(p)$ are the Zernike polynomials and the coefficients
$a_{lk}$ are obtained from the series expansion of $g_{l}(t)$
\begin{equation}
\label{e20}
g_{l}(t)=2\sum_{k=0}^{\infty}a_{lk}\sqrt{1-t^{2}}U_{l+2k}(t),
\end{equation}
where $U_{m}(t)$ are the Chebyshev polynomials of the second kind.
Since they are orthogonal in $[-1,1]$,
\begin{equation}
\label{e21}
a_{lk}=\frac{1}{\pi}\int^{1}_{-1}g_{l}(t)U_{l+2k}(t)dt.
\end{equation}

To use the symmetry properties most profitably, the best choice is
to measure $J(p_{z})$ for $p_{z}$ perpendicular to the [001]
direction. The directions of $p_{z}$ (Eq. (\ref{e18})) should be
equally spaced in the nonequivalent part of the BZ, which ranges
in the $\Gamma M K$ plane from $\varphi=0^\circ$ up to $45^\circ$
(or $30^\circ$) for cubic and tetragonal (or hexagonal)
structures, respectively. If $N$ projections are to be measured
for the structure with an $R$-fold rotation axis along the [001]
direction, the best orientation of the $n$-th projection
($n$=1,2,...,$N$) is given by the formula
$\varphi_n=\Delta\varphi/2+(n-1)\Delta\varphi$ where
$\Delta\varphi=\pi/(RN)$. However, if for any particular reason
one would like to perform measurements also for the main symmetry
directions, the orientation of the $n$-th projection ought to be
determined from the formula $\varphi_n=(n-1)\Delta\varphi'$ where
$\Delta\varphi'=\pi/[R(N-1)]$.

In this way
one can estimate $J(p_{z},p_{y})$, which represents the line integral
of the density along the [001] direction.
This procedure has already been applied to 1D ACAR spectra of yttrium
\cite{ref11}.
Even though the momentum density of Y is highly anisotropic,
only three projections were needed to obtain almost all details
of the 2D momentum density.
In this work, the method is succesfully applied to
beryllium (another hcp metal) in the extreme case of merely
two measured projections available.

\section{Application to beryllium data}

The Be Compton profiles were measured at the European Synchrotron
Radiation Facility (ESRF) beamline ID15B. The details of the
experiment are described in Ref.~\onlinecite{huotari2000a}. Two
projections were measured with $p_{z}$ along $\Gamma M$ and
$\Gamma K$ and with an overall momentum resolution of 0.1 a.u. The
corresponding highly accurate theoretical profiles were calculated
with the KKR-LDA method \cite{hamalainen1996}. Such
spectra allow the reconstruction of $\rho^{\mathbf L}\equiv
J(p_{z},p_{y})$ with $p_{x}$ along the hexagonal {\em c}-axis.
This 2D momentum density will be denoted as $\rho^{001}(p_z,p_y)$.

Let us choose the polar system $(p_{y}',p_{z}')$, fixed to the
lattice, with $p_{z}'$ along the $\Gamma K$ direction.
The measured profiles, $J(p_{z})\equiv J(t,\varphi)$, are described by the
polar angle $\varphi=0^{\circ}$ for $p_{z}$ along $\Gamma K$ and
$\varphi=30^{\circ}$ for $p_{z}$ along $\Gamma M$. In this case
two radial functions $g_{l}(t)$ with $l$=0 and 6 can be determined
from Eq.~(\ref{e18}),
\begin{eqnarray}
\label{e22} g_{0}(t) =& [&J_{\Gamma K}(t) + J_{\Gamma M}(t)]/2,
\nonumber
\\ g_{6}(t) =& [&J_{\Gamma K}(t) - J_{\Gamma M}(t)]/2.
\end{eqnarray}
It is convenient to choose the unit system so that
$t'=t/t_{max}=\cos(\alpha_i)$. The zeros of the
Chebyshev polynomials $U_{m}(t')$ occur then at
$\alpha_{i}=\frac{\pi}{2m}(2i+1)$ with $i=0,1,\ldots,m-1$, and
\begin{eqnarray}
\label{e23} a_{lk}=\frac{1}{M}\lbrace\sum_{i=1}^{M-1}g_{l}(\cos
(i\Delta\alpha)) \sin [(2k+l+1)i\Delta\alpha]\newline \nonumber
\\ + \frac{1}{2}(-1)^{k+l/2}g_{l}(0)\rbrace,
\end{eqnarray}
where $\Delta\alpha=\pi/2M$, and $M$ is the number of points used
in the evaluation of $a_{lk}$. In this work,
$M=720$ ($\Delta\alpha=0.125^\circ$), which allows one to
calculate 150 coefficients $a_{lk}$ for each $g_{l}(t)$.
This is equivalent to applying the Gaussian quadratures for the
polynomials up to the 300th order.

The FS of Be for the free-electron model is presented in Figs.~1
and 2.

\begin{figure}
\includegraphics[width=0.8\columnwidth]{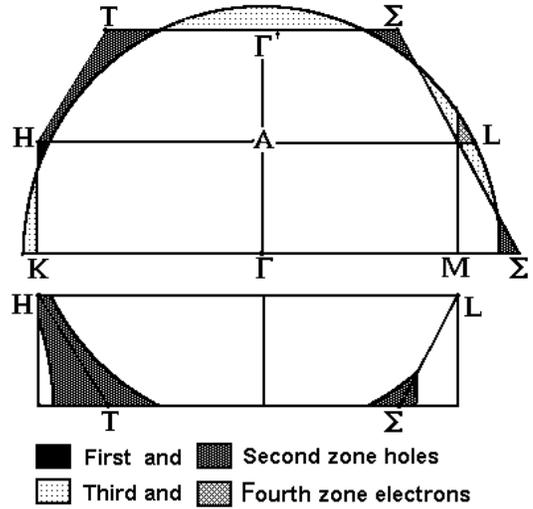}
\caption{Free-electron
Fermi surface of Be in the $\Gamma A K$ and $\Gamma A M$ planes in
the extended (upper panel) and reduced zone scheme (lower panel).}
\label{fig1}
\end{figure}
\vspace*{1\baselineskip}

To the authors' knowledge, the first realistic band-structure
calculations for Be were performed with the OPW method
\cite{ref14}. The results of that study, compared to the
free-electron model, suggested the following FS features: a) no
holes around the $H$ point either in the 1st or 2nd band (1st zone
fully occupied and no holes in the 2nd zone on the plane $AHL$);
b) very small holes around $\Sigma$ and reduced holes around $T$
in the 2nd zone when compared to the free-electron model; c) no
electrons around $\Gamma$ in the 3rd band; d) no electrons around
$L$ either in the 3rd or 4th bands; and e) cigars in the 3rd zone
around $K$ are larger than for the free-electron model with their
height close to $|KH|$ (see Fig. 2). These results were supported
by de Haas-van Alphen (dHvA) experiment \cite{ref15} with two main
differences: the computational cigars were triangular and the
coronet was slightly larger than seen by the dHvA experiment.

\begin{figure}
\includegraphics[width=0.8\columnwidth]{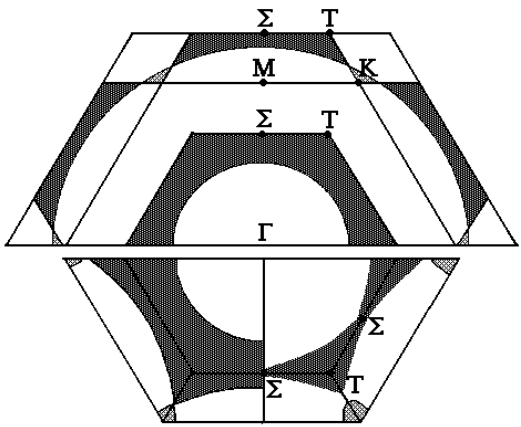}
\caption{The Fermi surface of Be in the $\Gamma M K$ 
plane. Upper panel: Free-electron model in the extended zone
scheme. Lower left panel: Free-electron model in the reduced zone
scheme. Lower right panel: Real FS derived from [14].}
\label{fig2}
\end{figure}
\vspace*{1\baselineskip}

Further theoretical band structure results \cite{ref16,ref17} were
similar to those in [14] while the latest LCAO results
\cite{ref18} turned out to be somewhat different,
i.e.~Fig.~3 in Ref.~\cite{ref18} representing $\rho({\bf p})$ along
$[011]\equiv \Gamma L$ points out that there are electrons around
the $L$ point in the 3rd and 4th bands. Theoretical Compton
spectra used in this work \cite{hamalainen1996} agree
qualitatively with the band structure results in
Ref.~\onlinecite{ref14}.

The results in the extended zone scheme along two main symmetry
directions $\Gamma K$ and $\Gamma M$ are displayed in Figs.~3 and
4. Fig.~3 presents the anisotropy of the 2D momentum density
$\rho^{001}({\bf p})$ for ${\bf p} \parallel \Gamma K$ and ${\bf
p}
\parallel \Gamma M$, reconstructed from both the experimental and
theoretical Compton profiles taken from
Refs.~\onlinecite{huotari2000a,hamalainen1996}. The widely-used
approximation to include the electron--electron (e--e) correlation
into the momentum densities is the so-called Lam-Platzman (LP)
correction \cite{lam-platzman74}. However, this correction is
isotropic and thus does not have any effect on the anisotropy of
theoretical curves in Fig.~3. The resolution function of the
experimental apparatus smears some of the fine structure of the
data as indicated in the figure.

\begin{figure}
\includegraphics[width=0.8\columnwidth]{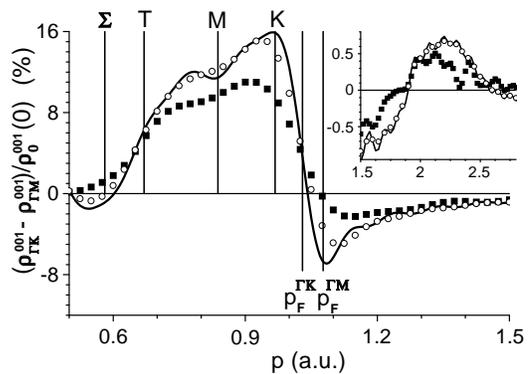}
\caption{Differences between 2D densities of Be for momenta along
$\Gamma K$ and $\Gamma M$, reconstructed from two Compton
profiles. The experimental results are indicated by solid squares.
The corresponding results of the reconstruction from theoretical
Compton profiles with and without resolution broadening are shown
by open circles and solid line, respectively.} \label{fig3}
\end{figure}
\vspace*{1\baselineskip}

For momenta $0.60<p<1.05$~a.u. (atomic units of momentum), the
reconstructed 2D densities have much larger values for
$p\parallel\Gamma K$ than for $p\parallel\Gamma M$. This is due to
the fact that for the real FS there are no electrons in the 3rd or
4th bands around the $L$ point. This is seen in Fig.~\ref{fig3} as
a positive anisotropy.  The largest positive values for $|\Gamma
M|\leq p\leq|\Gamma K|$~a.u.~are, additionally, a manifestation of
the lack of holes around the $H$ point. The negative anisotropy
for $p>1.05$ a.u., with the minimum at $p=p_{F}^{\Gamma M}$,
originates from the fact that $p^{\Gamma M}_F>p^{\Gamma K}_F$,
i.e.~the holes around the $\Sigma$ point, shown in the lower panel
in Fig.~2, are very small. Umklapp components for $p>1.9$ a.u.
along $\Gamma K$ are also observed. These results are in good
agreement with the theoretical 3D momentum density
\cite{hamalainen1996} and the experimental observation of the
Umklapp components \cite{huotari2002}. All subtle features of the
theoretical anisotropy are reproduced by the experiment in detail.
The accuracy of this result is due to the fact that even with
three Compton profiles available (the third one measured with
$\varphi=15^\circ$), the function $g_{6}(t)$ and thus also
$\rho^{001}_6(p)$ as well as the anisotropy, $\rho^{001}_{\Gamma
K}(p)-\rho^{001}_{\Gamma M}(p)=2\rho^{001}_{6}(p)$, would be the
same as obtained from two projections \cite{jura2001}. The fact
that the magnitude of the anisotropy is diminished indicates that
e--e correlation has an anisotropic effect on the momentum
density, as has been suggested earlier \cite{huotari2000a}.

Reconstructed densities for momenta along $\Gamma K$ and $\Gamma
M$ are presented in Fig.~4. It can be seen that the absolute
densities are not reproduced exactly, e.g.~a small electron-like
lens is observed at $p=0$~a.u. According to all previous
band-structure calculations, this element should not be present
and is thus most probably an artifact originating from the fact
that the isotropic component $\rho^{001}_{0}(p)$ is reconstructed
from only two projections. However, this artifact is cancelled out
in the difference shown in Fig.~\ref{fig3}, which further
demonstrates the accuracy of the obtained anisotropy.

\begin{figure}
\includegraphics[width=0.8\columnwidth]{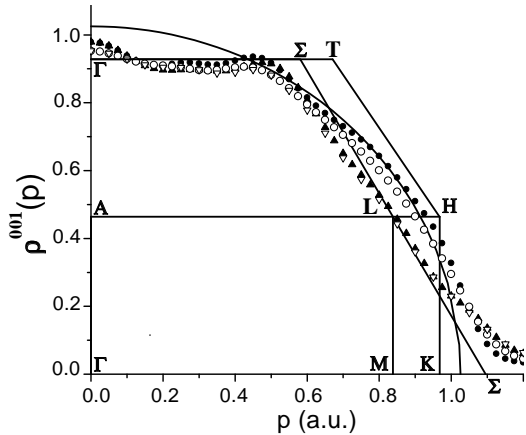}
\caption{2D momentum densities of Be for momenta along $\Gamma K$
(circles) and $\Gamma M$ (triangles), reconstructed from two
Compton profiles. Theory (including resolution-broadening and the
Lam-Platzman correction) and experiment are marked by solid and
open symbols, respectively.} \label{fig4}
\end{figure}
\vspace*{1\baselineskip}

In the analysis in the ${\bf p}$ space one should bear in mind
that $\rho^{001}(p_{z},p_{y})$ cannot be directly identified with
the FS line dimensions along the [001] direction because in real
metals $\rho({\bf p})<1$ also in the central FS. However,
undoubtedly the theoretical as well as the experimental results
show the lack of electrons around the $L$ point in the 3rd and 4th
bands and the shape of the FS on the $\Gamma MA$ plane close to
the double BZ boundaries. In order to obtain more detailed
information, $\rho^{001}(p_{z},p_{y})$ has to be folded to the
reduced zone scheme ({{\bf k} space}) as discussed in the
following.

\section{Results of the LCW folding procedure}

In the Compton-scattering experiment one measures the electron
density in the ${\bf p}$ space

\begin{equation}
\label{e27} \rho({\bf p}={\bf k}+{\bf G})=
\sum_{j}^{occ}\vert\int^{\infty}_{-\infty}\phi_{{\bf k}j}({\bf r})
e^{-i{\bf p}\cdot{\bf r}}d{\bf r}\vert^{2},
\end{equation}
where ${\bf k}$, ${\bf G}$, $j$ and $\phi_{{\bf k}j}({\bf r})$ denote
the wave vector, reciprocal lattice vector, band index and
electron wave function in the state ${\bf k}j$, respectively. The
summation is over all occupied states.

In a periodic lattice potential, the momentum density in the
$j^{\mbox{th}}$ band can be written as

\begin{equation}\label{e28}
\rho_{j}({\bf k}+{\bf G})=n({\bf k}j) ~ \vert u_{{\bf k}j}({\bf
G})\vert^{2},
\end{equation}
where $u_{{\bf k}j}$  are the Fourier coefficients of the electron
 wave functions and $n({\bf k}j)$ is the occupation
number.

It is well known that it is not possible to obtain the shape of
the FS knowing only the density $\rho({\bf p})$. This is due to
the fact that the density is not constant on the FS and $\rho({\bf
p})$ represents a sum of contributions from all occupied bands,
not only those crossing the Fermi energy. Thus, in order to obtain
the FS map, the best choice is to perform LCW folding \cite{ref9},
which is a conversion from ${\bf p}$ to ${\bf k}$ space. By doing
this, the folded densities visualize the shape of the individual
FS elements more clearly.

In the case of the electron density (neglecting correlation
effects), $\rho^{e}_{j}({\bf k})=\sum_{{\bf G}}\rho^{e}_{j}({\bf
k}+{\bf G}) =1$ (or 0 for unoccupied states) and  $\rho^{e}({\bf
k})=\sum_{j}\rho_j^{e}({\bf k})=n$, where $n$ denotes the number
of occupied bands contributing to the point ${\bf k}$. The
application of the LCW folding to $\rho^{\bf L}(p_{z},p_{y})$
gives the function $\rho^{\bf L}(k_{z},k_{y})$ which represents
the corresponding line integral of $\rho({\bf k})$, in this case
along ${\bf L}\equiv [001]$. The density $\rho^{\bf
L}(k_{z},k_{y})$ can be identified with the sum of
the FS line dimensions over occupied bands along {\bf L}.

The results of the folding procedure are displayed in Fig.~5,
where densities $\rho^{001}(k_{z},k_{y})$
reconstructed from two theoretical and two experimental Compton
profiles, are presented. The free-electron model result is also
shown for comparison.

Conventionally, the momentum densities $\rho({\bf k})$ obtained
via the LCW folding, and thus also densities
$\rho^{001}(k_{z},k_{y})$, are presented in arbitrary scale. As
was shown by Kaiser et al. \onlinecite{Kaiser}, the main reason is
that the convergence of the LCW procedure is dependent on the
degree of the localization of the electrons. The more localized
they are, the larger the contribution of the Umklapp components at
high momenta is, which consequently increases the truncation error
of the LCW procedure. This effect decreases the LCW-folded
densities in a nonuniform way, and thus an absolute normalization
is difficult. In the case of experimental densities, the finite
resolution function and statistical noise, both affecting
especially the Umklapp components, complicate the normalization
even further.

In Be the volume of occupied valence states is equal to the volume
of two BZ's. Thus, if two bands were completely filled,
$\rho^{001}(k_{z},k_{y})$ would be constant and equal to $2|\Gamma
A|$. The light and dark areas in Fig.~5 depict low and high
electron densities originated from the holes in the second band
(coronet) and the electrons around $K$ in the third band (cigars),
respectively.

\begin{figure}
\includegraphics[width=0.8\columnwidth]{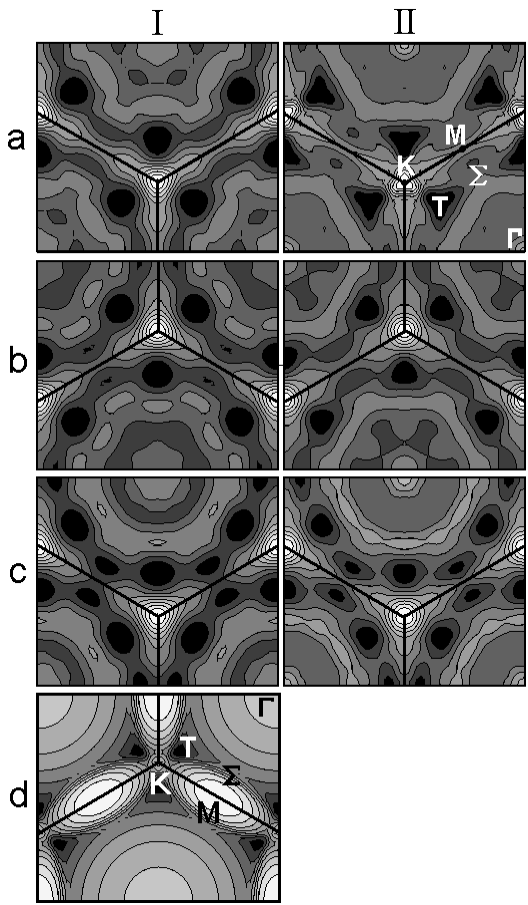}
\caption{2D LCW densities $\rho^{001}(k_{z},k_{y})$ in Be, in the
repeated zone scheme, reconstructed from two 1D projections.
Panel (a): theory without LP correction, panel (b): theory with
LP correction and resolution broadening, panel (c): experiment, and
panel (d): free-electron model.
Columns I and II show
results described by 60 and 90 (or 120 in panel (a)) orthogonal polynomials, respectively.
Each figure contains 10 contourlines.} \label{fig5}
\end{figure}
\vspace*{1\baselineskip}

Fig.~6 shows the same densities in more detail along the main
symmetry directions together with densities obtained from the
corresponding LP-corrected theoretical Compton profiles. From
Figs.~5 and 6 the 2nd-zone holes around $T$ and the 3rd-zone
electrons around $K$ can be clearly recognized. Other features,
e.g.~the electron-like artifact around the $\Gamma$ point in the
3rd band (seen also in Fig.~4) and a hole-like artifact between
the $\Sigma$ and $M$ points are caused by using only two
components of reconstructed density $\rho^{001}(p_{z},p_{y})$,
determined from only two Compton profiles.

\begin{figure}
\includegraphics[width=0.8\columnwidth]{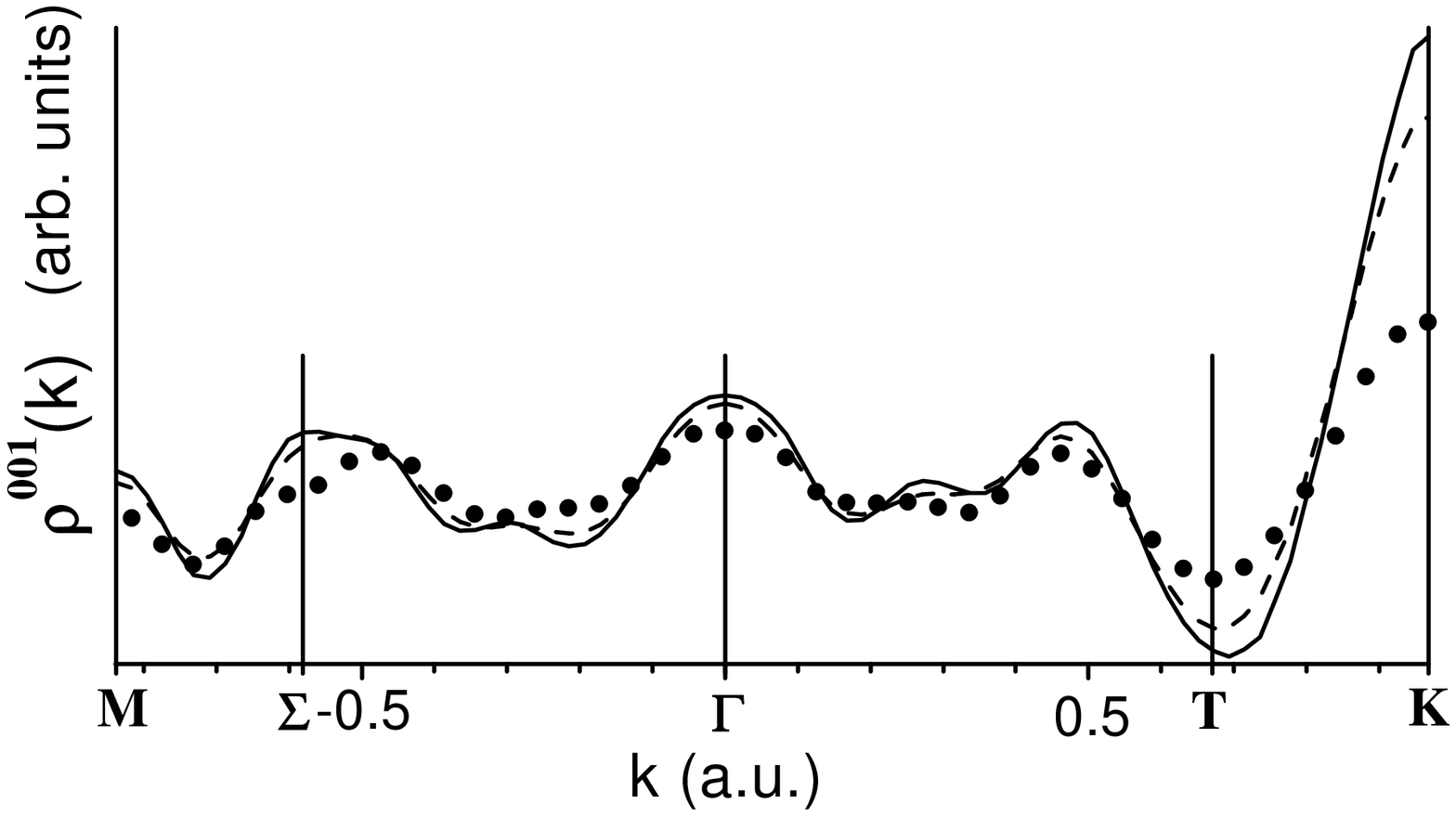}
\caption{2D LCW densities in Be for momenta along $\Gamma K$
(right-hand side) and $\Gamma M$ (left-hand side), reconstructed
from two 1D spectra. The theory with LP correction with and
without resolution broadening is indicated by solid and dashed
lines, respectively, while the experiment is marked by circles.}
\label{fig6}
\end{figure}
\vspace*{1\baselineskip}

\section{Summary}

When the electron momentum density $\rho({\bf p})$ is highly
anisotropic, it is quite difficult to obtain its shape from 1D
projections accurately, since a relatively large number of
projections is needed. For these cases, the
reconstruction of 2D momentum densities $J(p_{z},p_{y})$ instead
of 3D densities $\rho({\bf p})$ is advised. It is shown that only a small
number (2-5) of measured plane projections is needed to
reconstruct the 2D momentum density, and generally this 2D density
allows much more detailed analysis of the FS than 1D projections.
The use of the Cormack method is proposed, because a) the
expansion of measured spectra into orthogonal polynomials has
mean-squares approximation properties and thus it effectively
smoothes the statistical noise in the experimental data
\cite{Bjorck}; b) the method employs the Chebyshev
polynomials, which are the only orthogonal polynomials having
analytical zeros, allowing the estimation of the coefficients
$a_{lk}$ to a high degree of precision.

This method was applied to two Be Compton profiles and the line
projection of the momentum density along the [001] direction was
reconstructed.

The analysis performed in the extended ${\bf p}$ space shows that
for  $0.60<p<1.05$ a.u.~the line dimensions of the FS along [001]
are much larger for $p\parallel\Gamma K$ than for
$p\parallel\Gamma M$, which is caused by the lack of electrons in
the 3rd and 4th bands around the $L$ point and the lack of holes
around the $H$ point. The anisotropy of the Fermi momentum,
i.e.~$p^{\Gamma M}_F>p^{\Gamma K}_F$, leading to very
small holes around the $\Sigma$ point, is also observed. All
subtle features of the theoretical anisotropy are reproduced by
the experiment in detail. However, its magnitude is diminished,
which indicates that e--e correlation has an anisotropic effect on
the momentum density. With the example of the anisotropy
$\rho^{001}_{\Gamma K}(p)-\rho^{001}_{\Gamma M}(p)$, it is shown
how symmetry properties can enhance the accuracy of the
reconstruction.

After folding the reconstructed densities
$\rho^{001}(p_{z},p_{y})$ into the reduced zone scheme,
the 2nd zone holes and 3rd zone
electrons could be clearly recognized
around the $T$ and $K$ points, respectively. Moreover,
the results both in {\bf p} and {\bf k} spaces clearly show the
absence of the FS elements around the $L$ point in the 3rd and 4th
bands.

\acknowledgments

The authors are grateful to S.~Kaprzyk and A.~Bansil for making
the theoretical Compton profiles of Be available.
The European Synchrotron Radiation Facility is acknowledged
for the provision of synchrotron radiation facitilities and the
authors would
like to thank Dr.~Thomas Buslaps for assistance in using beamline ID15B.
M.~S.-C.~would like to thank CELTAM (Centre for Low
Temperature studies of Promising Materials for Applications) in
Poland and the Academy of Finland for financing her research visit
at the University of Helsinki. S.H., K.H.~and S.M.~were supported by the
Academy of Finland (contracts No.~201291/40732). S.H.~was
supported also by the National Graduate School in Material Physics,
which is funded by the Finnish Ministry of Education.


\end{document}